\documentclass{aa}

\usepackage{graphicx}
\usepackage{amsmath}
\usepackage[varg]{txfonts}
\begin{document}
   \title{ALMA detection of the rotating molecular disk wind from the young star HD 163296}
\titlerunning{Rotating disk wind from HD163296}

   \author{P.~D. Klaassen          \inst{1}\thanks{email: klaassen@strw.leidenuniv.nl},
A Juhasz\inst{1}
G.S. Mathews\inst{1}, 
J.C. Mottram\inst{1},
I. De Gregorio-Monsalvo\inst{2,3},
E.F. van Dishoeck\inst{1,4},
S. Takahashi\inst{5}, 
E. Akiyama\inst{6},
E. Chapillon\inst{5}, 
D. Espada\inst{6}, 
A. Hales\inst{2}, 
M.R. Hogerheijde\inst{1},
M. Rawlings\inst{7}, 
M. Schmalzl\inst{1},
L. Testi\inst{3,8}
	          }
\authorrunning{Klaassen et al.}
   \institute{Leiden Observatory, Leiden University, P.O. Box 9513, 2300 RA, Leiden, The Netherlands
\and
Joint ALMA Observatory, Alonso de Cordova 3107, Vitacura, Santiago, Chile
\and
European Southern Observatory, Karl Schwarzschild Str 2, 85748, Garching, Germany
\and
Max-Planck-Institut f\"ur Extraterrestrische Physik, Giessenbachstrasse 1, 85748 Garching, Germany
\and
Academia Sinica Institute of Astronomy and Astrophysics, P.O. Box 23-141, Taipei 10617, Taiwan 
\and
National Astronomical Observatory of Japan (NAOJ), 2-21-1 Osawa, Mitaka, Tokyo 181-8588, Japan
\and
NRAO, 520 Edgemont Road, Charlottesville, VA 22903, USA
\and
INAF - Osservatorio Astrofisico di Arcetri, Largo E. Fermi 5, 50125, Firenze, Italy
}
   \date{}

  \abstract
  {Disk winds have been postulated as a mechanism for angular momentum release in protostellar systems for decades.   HD 163296 is a Herbig Ae star surrounded by a disk and has been shown to host a series of HH knots (HH 409) with bow shocks associated with the farthest knots. Here we present ALMA Science Verification data of CO $J$=2-1 and $J$=3-2 emission which are spatially coincident with the blue shifted jet of HH knots, and offset from the disk by -18.6 km s$^{-1}$. The emission has a double corkscrew morphology and extends more than 10$''$ from the disk with embedded emission clumps coincident with jet knots. We interpret this double corkscrew as emission from material in a molecular disk wind, and  that the compact emission near the jet knots is being heated by the  jet which is moving at much higher velocities. We show that the $J$=3-2 emission is likely heavily filtered by the interferometer, but the $J$=2-1 emission suffers less due to the larger beam and  sensitivity to larger scale structures.  Excitation analysis suggests temperatures exceeding 900 K in these compact features, with the wind mass, momentum and energy being of order 10$^{-5}$ M$_\odot$, 10$^{-4}$ M$_\odot$ km s$^{-1}$ and 10$^{40}$ erg respectively. The high mass loss rate suggests that this star is dispersing the disk faster than it is funneling mass onto the star.  }


   \maketitle
%
%________________________________________________________________

\section{Introduction}

Disks are crucial for the formation of stars because they are a mass reservoir for building up the star and any planets forming in the system.  They mediate accretion onto the star, and both jets and winds from the disk act as release mechanisms for angular momentum during accretion. Jets and winds are generally seen quite early on in the star formation process \citep[e.g Class 0 sources,][]{Arce07}, but have also been observed in the final stages of the formation process as well \citep[e.g. TW Hydra,][]{Pascucci11}.  For a complete theory of star formation, we need to understand the mechanisms responsible for jet and wind launching. We need to understand how much of the disk material ends up on the star, and how much is removed from the system to constrain how efficiently the star is gaining mass. There is already evidence for residual rotation in some outflows launched from disks \citep[e.g.][]{launhardt09}, and when the kinematics of the jet and disk are both measured, their rotations appear to be consistent with each other \citep[e.g. DG Tau,][]{Testi02,Bacciotti02}, showing the causal links between these phenomena. HD 163296 is a good candidate for studying this phenomena because of the favourable viewing angle and the presence of the HH 409 system.

HD 163296 is an isolated, prototypical Herbig Ae star which shows  an infrared excess and is surrounded by a Keplerian disk \citep{Isella07,Hughes08}. It is at a distance of 122 pc, and has an age of 4 Myr \citep{vandenancker98}. The scattered light observations of \citet{Wisniewski08} suggest time variability in the disk, which they attribute to inflation of the inner disk edge, consistent with the infrared variability described in \citet{Sitko08}.  These authors also show evidence for an emission flare at 3 $\mu$m in 2002, which they suggest may have been due to a combination of inner disk flare-up and material being released from the disk into the surroundings.  Using FUV data, \citet{Deleuil05} derived a mass outflow rate of \mbox{$7\times10^{-9}$} \mbox{M$_\odot$ yr$^{-1}$} from the star, while  \citet{Hubrig09} and \citet{Garcia06} find  mass accretion rates of $\sim10^{-7}$ M$_\odot$ yr$^{-1}$. \citet{Kraus08} presented VLT/AMBER resolved Br$\gamma$ observations of HD 163296 and suggest that the extended Br$\gamma$ emission comes from a stellar wind or disk wind.  The NIRSPEC observations of \citet{Salyk11} show rovibrational CO emission knots at approximately $\pm$ 100 km s$^{-1}$ at a velocity resolution of 12.5 km s$^{-1}$ which is not included  in their disk model, but may be evidence of a high velocity wind.

HD 163296 is also associated with Herbig Haro (HH) object 409 \citep{Devine00}. HH 409 is a chain of HH knots traced by  [S{\sc ii}] and H$\alpha$ seen in both a jet and counter jet which is oriented perpendicular to the disk plane  \citep[e.g.][]{Wassell06}. \citet{Sitko08} suggested that these knots are ejected on 5-6 yr intervals.    The PSF subtracted coronographic images  of the region by \citet{Wassell06} show a series of HH knots which they labelled A through D.  The jet is blue shifted, and the knots are labelled A and A2. The counter-jet is redshifted, and the knots are labelled B-D. For the brightest components (A in the jet, and C in the counter jet), multi-epoch observations were used to derive proper motions of $0.\!''49$ yr$^{-1}$ for knot A, and 0.\!$''$35 yr$^{-1}$ for C, with jet velocities greater than 250 km s$^{-1}$. No sub-mm molecular outflow counterpart to this jet has been imaged before  however there may be evidence for a blue shifted wind in the spectra of  \citet{Thi04}.

To clarify our general terminology, we refer to a jet as high velocity gas ($\gtrsim$ 100 km s$^{-1}$) launched from the inner 0.1 au of the disk,  collimated by the magnetic field \citep[e.g.][]{Ray07,Reipurth01}. Jets are generally observed in optical/near-infrared atomic lines, and bow shocks resulting from the interaction of the jet with cloud material are often observed as HH objects \citep[e.g.][]{Reipurth01}.  We refer to outflows as the molecular gas entrained by jets and/or winds as they push through the ambient material \citep[e.g.][]{Arce07}.  These flows are generally observed to have much lower velocities ($<$ 50 km s$^{-1}$) because they comprise ambient material which has been swept up by the jet \citep{Koenigl00}. Within this framework, we define disk winds as the low energy material launched from the disk at radii of a few au \citep[e.g.][]{Pudritz07}.  A disk wind ($<$ 25 km s$^{-1}$) is a centrifugally driven, magnetically collimated wind blown off the surface of the disk {\citep{PN83}. It is distinct from an outflow, and can indeed entrain an outflow of its own.} An overview of these definitions is given in \citet{Mckee07}.

Here we present ALMA Science Verification data of two CO transitions, $J$=2-1 (Band 6, 230.538 GHz) and $J$=3-2 (Band 7, 345.796 GHz) of the disk wind associated with HD 163296 and HH 409.  The disk gas emission is presented separately (Mathews et al., de Gregorio-Monsalvo et al., submitted, and Chapillon et al., in prep) and will not be discussed here.
In Section \ref{sec:obs}, we summarize our observations and data reduction.  In Section \ref{sec:results} we describe the morphology and energetics of the observed disk wind.  In Section \ref{sec:discussion} we relate these findings to the disk and jet properties previously determined for this source, and summarize in Section \ref{sec:conclusions}.

\section{Observations and Data Reduction}
\label{sec:obs}

These $^{12}$CO $J$=2-1 and $J$=3-2 observations were taken with ALMA as part of the Science Verification program (2011.0.000010.SV).  Both datasets were observed in June and July 2012 with 18 - 20 antennas in an extended configuration ($\sim$20-400 m baselines).  The Band 6 data consisted of three separate observing blocks, and the Band 7 data are a combination of five. The total time on source in each band was 46 and 59 minutes, respectively.  The quasars J1924-292 and J1733-130 were used for bandpass and phase calibration. For the Band 6 data, Juno, Neptune and Mars were used for amplitude calibration. For four of the Band 7 executions, Neptune was observed for amplitude calibration, and this amplitude calibration was applied (bootstrapped) to the fifth execution for which Juno was observed, but was too close to the horizon to be properly used for amplitude calibration. Data reduction and imaging were all performed using the Common Astronomy Software Applications \citep[CASA,][]{CASA}. Both data sets were cleaned using natural weighting, and the channel spacings of the spectral windows for the CO lines were 244 kHz and 122 kHz, which were resampled to spectral resolutions of 0.32 and 0.25 km s$^{-1}$ in Bands 6 and 7 respectively.  Both datasets were self-calibrated in both phase and amplitude using line-free continuum channels in all four spectral windows.The per-channel rms noises in these two datasets are  6.9 and 4.2 mJy beam$^{-1}$ after self-calibration.  Data from two antennas (PM01 and DV03) were removed for the July observations in Band 6 because of anomalously high amplitudes. HCO$^+$ $J$=4-3 was simultaneously observed with CO $J$=3-2 in the Band 7 observations, and thus has the same noise properties. These data are presented in Mathews et al. (Submitted).

\begin{figure}
\includegraphics[width=\columnwidth]{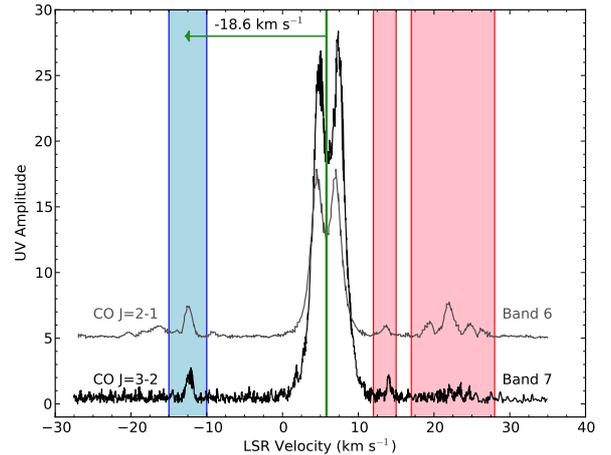}
\caption{Visibility spectra averaging over the two shortest baselines in each of the Band 6 and Band 7 datasets showing the blue and red shifted wind emission in this field of view. The rest velocity of the source is 5.8 km s$^{-1}$ \citep{Hughes08} as shown with a green line in the plot. The Band 6 (CO $J$=2-1) data have been shifted upwards by 5 to clearly show both spectra.}
\label{fig:UV_spectrum}
\end{figure}

The synthesized beams for the two sets of observations are shown in Table \ref{tab:limits}. The shortest baselines in the array dictate the largest scale observable in each dataset, and the formula for determining the approximate largest angular  scales (LAS) is given in Section 6.2 of the ALMA Cycle 1 Technical Handbook\footnote{https://almascience.nrao.edu/documents-and-tools/cycle-1/alma-technical-handbook}. The largest scale structures to which our observations are sensitive are given in Table \ref{tab:limits}. Any structures larger than this are filtered out by the interferometer. 

\begin{table}
 \caption{Properties of the 230 GHz and 345 GHz Observations.   The rms noise values listed below have units of mJy beam$^{-1}$ per channel. The channel spacings are 0.32 and 0.25 km s$^{-1}$ for the 230 and 345 GHz data (respectively)}
\begin{tabular}{lllll}
\hline \hline
Line   & rms& Synthesized & Primary &Largest\\
& &Beam, P.A.  & Beam & Observable Scale\\
 \hline
 $J$=2-1 & 6.9& 0.80$''\times$0.68$''$, 75$^\circ$ & 26$''$ & 11$''$\\
 $J$=3-2 & 4.2& 0.68$''\times$0.44$''$, 91$^\circ$ & 17$''$& 8$''$\\
 \hline
 \end{tabular}
  \label{tab:limits}
 \end{table}

\section{Results}
\label{sec:results}

\begin{figure*}[ht]
\includegraphics[width=\textwidth]{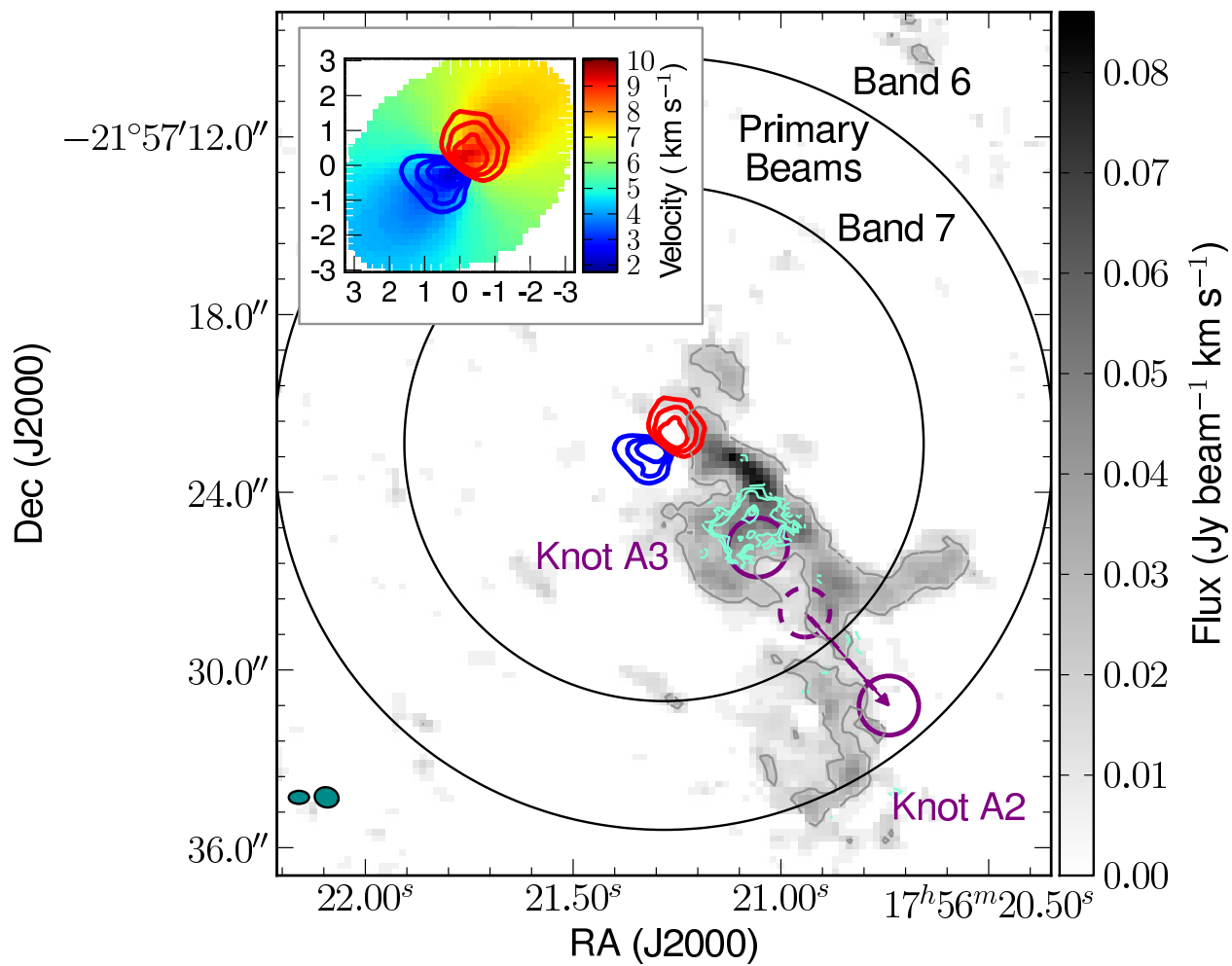}
\caption{Blue shifted disk wind from HD 163296. The greyscale and single (4$\sigma$) contour show the CO $J$=2-1 integrated intensity of the blue shifted wind from HD 163296. The blue and red contours show the 15, 20 and 25 times the  rms noise (20 mJy beam$^{-1}$)  of the blue (-2 to 5.5 kms$^{-1}$) and red (6 to 12 km s$^{-1}$)  HCO$^+$ $J$=4-3 emission in the disk. The inset image shows the first moment map of the HCO$^+$ $J$=4-3 emission in the disk, with the same red and blue contours as in the main figure for reference. The two black arcs show the primary beams of the 345 GHz (smaller) and 230 GHz (larger) observations. The cyan contours show 4, 6, and 8 times the rms noise in the CO $J$=3-2 emission (2.6 mJy beam$^{-1}$).  The two solid purple circles show the 2012 positions of knots A2 and A3 as described in the text. The dashed purple circle and arrow show the position of knot A2 in 2004, and its projected motion to 2012. The two dark cyan circles in the bottom left corner show the synthesized beams of the Band 7 (left) and Band 6 (right) datasets. The dimensions of the synthesized beams are given in Table \ref{tab:limits}. }
\label{fig:CO_21_mom0}
\end{figure*}

Figure \ref{fig:UV_spectrum} shows the averaged visibility spectra of  the two shortest baselines (21.2 and 21.7 m) in both datasets.  The velocity integration limits used for finding emission in the image plane are shown as red and blue boxes in Figure \ref{fig:UV_spectrum}. This wind emission is at distinct velocities from the disk itself (centered at the green line in Figure \ref{fig:UV_spectrum}).   The redshifted emission is strongest in the Band 6 data. At both the red and blue shifted velocities highlighted in Figure \ref{fig:UV_spectrum}, there is phase coherence, and it is offset from zero.  This suggests that the emission is coming from a region offset from the observing center.

We detect blue shifted wind emission in the image plane in both CO $J$=2-1 and $J$=3-2.   Figure \ref{fig:CO_21_mom0} shows the integrated intensities of the CO emission as traced by CO $J$=2-1 in the greyscale and in CO $J$=3-2 in cyan contours.  The red and blue contours in the center of the figure and in the inset show the 15, 20, and 25 $\sigma$ contours of the integrated blue and red shifted HCO$^+$ $J$=4-3 emission in the disk. This is shown to give context to the position of the wind. We have plotted (using dashed lines) the 2004 position of knot A2 and its proper motion. Also shown (with  solid purple circles) are the expected positions of knots A2 and A3 in 2012 using the proper motion of knot A given in \citet{Wassell06}. This correction for proper motion is described more fully in Section \ref{sec:discussion}.   The pointing center is $\alpha$=17:56:21.28, $\delta$=-21:57:22.36, and the large black circles show the full width at half maximum power of the Band 6 and 7 primary beams (the FWHM of the Band 6 primary beam is larger than the Band 7). Neither of the datasets shown in Figures \ref{fig:CO_21_mom0} and \ref{fig:CO_21_mom1} are primary beam corrected.   The integrated and peak intensities of the CO $J$=2-1 emission are 2.58 Jy km s$^{-1}$ and 0.084 Jy beam$^{-1}$ km s$^{-1}$, respectively.  The disk wind traced by CO $J$=2-1 is a continuous structure with what appears to be a forked end. From the first moment maps of both transitions (the CO $J$=2-1 first moment map is shown in Figure \ref{fig:CO_21_mom1}), we find that both transitions show velocity gradients perpendicular to the wind and jet direction, consistent with that in the disk (blue-shifted to the south-east, red-shifted to the north-west).

The spatial extent of the $J$=3-2 emission is much smaller than the $J$=2-1 emission (see Figure \ref{fig:CO_21_mom0}).  It is concentrated in two knots; one within the full width at half maximum of the primary beam, and one external to it. The brightest CO $J$=3-2 knot is well within the primary beam, and has integrated and   peak intensities of 0.6 Jy km s$^{-1}$ 0.06 Jy beam$^{-1}$ km s$^{-1}$ respectively. We suggest it is associated with a jet knot ejected in 2002.  The second knot(A2), outside the Band 7 primary beam, is coincident with where HH knot A2 should be in 2012 using the proper motions derived in \citet{Wassell06}. This knot is best seen when the Band 7 data is primary beam corrected (as shown in Figure \ref{fig:filtering}). Primary beam correction takes into account the attenuation of the signal away from the pointing center.  The noise in the map is no longer uniform, but increases with distance from center.  The smaller and larger arcs in Figure \ref{fig:filtering} show where the Band 7 and Band 6 primary beams are attenuated to 50\%.  The primary beam corrected data extends to where the primary beam is attenuated to 20\% of its sensitivity at the pointing center.  The CO $J$=3-2 emission in the vicinity of HH knot A2 is much more prominent in this representation.

The differing morphology between the two CO transitions may be explained by spatial filtering by the interferometer. The Band 7 observations cannot detect structures larger than about 8$''$, while the Band 6 observations are sensitive to larger scale structures ($\sim$ 11$''$).  It is a combination of the higher excitation conditions required for the $J$=3-2 transition and the spatial filtering which is causing the wind to appear clumpy and compact in the Band 7 observations. We masked the shortest baselines in the Band 6 data to match the largest angular scales recoverable in the Band 7 data (8$''$ at 230.5 GHz corresponds to 16.6 k$\lambda$).  In this filtered data, we find that the wind, as traced by CO $J$=2-1, appears clumpier, and peaks close to the $J$=3-2 emission.  That the regions with the strongest emission for two species are slightly different suggests that the excitation of the two lines also plays a role in the morphology.

In the image plane (see Figure \ref{fig:CO_21_mom0}), we do not detect the northern redshifted gas.  It is likely that the redshifted wind is filtered out by the interferometer.  This is evidenced by the redshifted emission being detected in the Band 6 visibilities, but not the Band 7 visibilities. The redshifted emission is filtered out on all but the shortest two baselines (at $\sim$ 21 m). The next shortest baseline is 27.5 m, and the short baseline sampling appears to be too coarse to image the red lobe. Followup observations covering a wider areas, and recovering larger scale structures are required to confirm this hypothesis.

\section{Discussion}
\label{sec:discussion}

\citet{Wassell06} characterized the knots of HH 409 associated with HD 163296. For knot A they suggest a proper motion of $0.\!''49$ yr$^{-1}$. Using that proper motion, and extrapolating to 2012, we find that knot A is well beyond the primary beam of our observations.  However, if we apply the proper motion of knot A to knot A2, we find that it should have been approximately 11$''$ from the disk when our observations were taken. Using a position angle of 42$^\circ$ \citep[taken from][]{Wassell06}, this knot should be slightly ahead of the CO emission seen in the primary beam corrected image of Figure \ref{fig:filtering} as shown with a purple circle.  Note that this clump, at 11$''$, is in the region where the interferometer is only 20\% as sensitive as it is at the pointing center. That a jet knot should be co-located with this CO $J$=3-2 emission feature suggests that the molecular emission is real despite being far from the pointing center.

The observations of \citet{Sitko08} showed a flare in 3 $\mu$m emission coming from HD 163296 in 2002.  If we assume that this was a knot launching event, and assume this knot has the same proper motion as HH knot A ($0.\!''49$ yr$^{-1}$), then, at the epoch of our observations (10 years later), the ejecta should be 4.9$''$ from the disk. This is the location of  the leading edge of our strongest CO $J$=3-2 knot, which we named A3 (see Figure \ref{fig:CO_21_mom0}). It should be noted that an atomic HH knot from this flare would have been obscured by the coronograph of the 2004 observations of \citet{Wassell06}, but that subsequent observations should show the atomic knot emission \citep[see for instance,][and Ellerbroek et al. in prep.]{Gunther13}.

\begin{figure}
\includegraphics[width=\columnwidth]{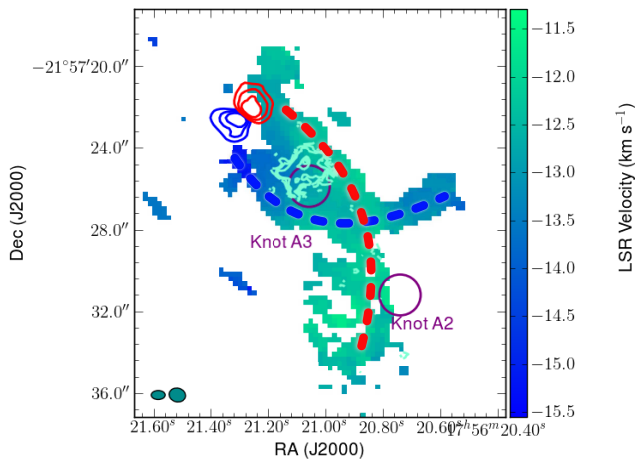}
\caption{CO $J$=2-1 intensity weighted velocity map of the blue shifted wind from HD 163296. The purple circles show the 2012 positions of the knots, the blue and red contours show the HCO$^+$ $J$=4-3 emission, and the two dark cyan circles in the bottom left corner show the synthesized beams of the observations, as in Figure \ref{fig:CO_21_mom0}.  The blue and red dashed lines delineate the double corkscrew described in the text. }
\label{fig:CO_21_mom1}
\end{figure}

\begin{figure}
\includegraphics[width=1.05\columnwidth]{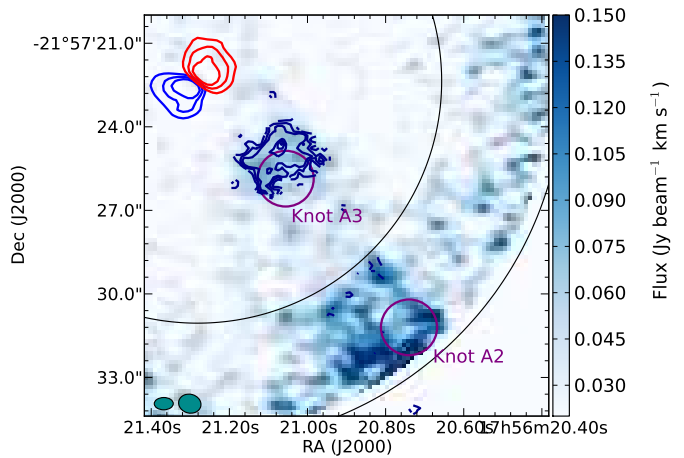}
\caption{CO $J$=3-2 integrated intensity map of the blue shifted wind from HD 163296. The colourscale shows the primary beam corrected integrated intensity of the $J$=3-2 emission  The black, blue and red circles and contours are the same as Figure \ref{fig:CO_21_mom0}.  The molecular counterpart to HH knot A2 is more prominent in this image because we have corrected for the fall-off in sensitivity as a function of distance from the pointing center.   }
\label{fig:filtering}
\end{figure}

Based on the emission morphologies shown in Figures \ref{fig:CO_21_mom0} and \ref{fig:CO_21_mom1}, we suggest that the $^{12}$CO emission is primarily tracing the disk wind coming from HD 163296. The morphology of the emission, and its velocity structure suggests that this is a disk wind, and not an entrained outflow from the jet.  That there appears to be two arms to the wind argues against entrainment because it is unlikely that a single knot can produce two velocity components (like those highlighted by dashed lines in Figure \ref{fig:CO_21_mom1}).  If the CO were entrained material, there is no explanation for the highly blue shifted material at the end of the bluest arm of the observable wind,  since there is no HH knot ahead of it which could have entrained that gas.  In the area around the knots, the molecular gas in the wind is being heated by the jet; either from the warm shocks, or directly from the jet heating the dust. Because the temperature is elevated, smaller amounts of gas appear brighter.  This means that compact  and bright knots of CO will appear in the larger scale wind. In the case of CO $J$=3-2, the larger scale wind is resolved out by the interferometer, but the bright and compact knots of heated gas are still detectable. The CO $J$=2-1 emission is also quite bright in this region.  Note that the enhanced CO emission is slightly closer to the disk ( i.e. upstream) than the expected position of the knots in 2012, reinforcing the suggestion that this material was heated by a likely HH knot.  

The velocity structure shown in the CO $J$=2-1 first moment map (Figure \ref{fig:CO_21_mom1}) suggests that there may be two foot points for the disk wind, which are on opposite sides of the disk, co-rotating with it.  The velocity field is an imprint of the rotation of the disk, and the wind itself may be collimated by the magnetic field. The bluest portion of the disk wind is emitting from the south east part of the disk, which itself is known to be blue shifted. The reddest part of the disk wind is coming from the north west portion of the disk, which is red shifted. Some jets do show evidence for rotation like that seen here  \citep[e.g. DG Tau][]{Bacciotti02}. We suggest this disk wind has a double corkscrew, similar to the modelled disk wind phenomena seen  in, for instance, Figure 3 of \citet{Ouyed03} or Figure 7 of \citet{launhardt09}, but with two components. The dashed blue and red lines in Figure \ref{fig:CO_21_mom1} give an illustration of the double corkscrew described here. The two components of the wind cross each other along the line of sight  just further away from the disk than the A3 knot seen in CO $J$=3-2.  This double corkscrew can explain why the outer edges of the $J$=2-1 emission appear forked, and with very different velocities; they are at opposite ends of the double corkscrew.

 The recent observations of \citet{Gunther13} show Ly$\alpha$ emission associated with the HH knots, which indicates that the gas in these regions is hot. Using RADEX, a non-LTE 1D radiative transfer calculator \citep{vandertak07}, we determined the excitation temperature and column density necessary to produce the ratio of  line peak intensities of the $J$=2-1 (1.195 K)  and $J$=3-2 (1.523 K) transitions of CO integrated over the disk wind emission near knot A3.  For consistency, we used the spatially filtered Band 6 data for comparison with the Band 7 data.   Using an ambient density of 1.9$\times10^{3}$ cm$^{-3}$ \citep[see Table 8 of ][]{Wassell06}, we find that to match our line ratio requires a column density of 1$\times10^{15}$ cm$^{-2}$.    Dividing this derived column density by the assumed volume density recovers our observed emitting region when we make the assumption of spherical symmetry. For higher column densities, the line ratio does not converge to the observed $J$=2-1/$J$=3-2 ratio of 0.8. For lower column densities, the line peak intensities are both below 1 K. At this column density the line ratio approaches 0.8 for temperatures greater than 900 K. Above 1000 K, the $J$=2-1 emission has a negative excitation temperature.  We can recover our line ratio and integrated intensities at lower temperature, however this results in a radius ten times greater than that observed when dividing the column density by the volume density. We therefore use an excitation temperature of 960 for further calculations.  At these densities and temperatures, both lines are optically thin, with opacities near 0.1. These high temperatures are not unrealistic, especially since the region we are probing is being heated by a jet knot.

Using the derived column density and assumed density, the depth of the emission region towards the brightest CO$J$=3-2 knot ($\sim$ 300 au) is roughly the same as the width of the emission. This is consistent with the flow being cylindrical. To calculate the mass and energetics of the outflow, we used the integrated intensity (moment zero, e.g. Figure \ref{fig:CO_21_mom0}) and intensity weighted velocity (moment one, e.g. Figure \ref{fig:CO_21_mom1}) maps of the CO emission in both CO transitions.  When creating these maps, the data were clipped at 3.5 and 4 $\sigma$ in each channel when creating the zeroth and first moment maps (respectively).   We assume a wind age of 400 yr based on the average velocity in the wind (-18.6 km s$^{-1}$) and its spatial extent in the Band 6 observations ($\sim 13''$). This will underestimate the age of the wind if it is more extended than the Band 6 primary beam. This is the age used to determine the wind forces, mechanical luminosities and mass loss rates quoted in Table \ref{tab:energy}.

To calculate the mass, we converted the integrated intensity of the line into a CO column density assuming local thermodynamical equilibrium, a temperature of 960 K and a Jansky to Kelvin conversion based on the synthesized beams of the Band 6 (43.48 Jy/K) and Band 7 (33.78 Jy/K) data.   This was converted to a total molecular column density using a CO abundance of 10$^{-4}$ with respect to molecular hydrogen, and a mean molecular mass of 2.8 \citep[see, for instance, appendix A of ][]{Kauffmann08}. The wind momentum was derived by determining the mean velocity of the gas (-18.6 km s$^{-1}$), and multiplying that by the derived mass.  We calculated the wind force by dividing the momentum by the wind age. Similarly, the mechanical energy in the wind was calculated using E= $\frac{1}{2} mv^2$, and the mechanical luminosity was determined by dividing the energy by the assumed age of the outflow.  Using the column density derived from the RADEX calculation, the outflow mass drops by almost an order of magnitude.

 The angular momentum ($L=mvr$) of the A3 knot is 3.7$\times10^{-6}$ M$_\odot$ km$^2$ s$^{-1}$, where $r$ is the radius of the CO $J$=3-2 emission ($r=133$ au).  A mass accretion rate of  $\sim10^{-7}$ M$_\odot$ yr$^{-1}$ was previously determined for this source  \citep{Garcia06,Hubrig09}. This is a factor of two lower than the mass loss rate we present in Table \ref{tab:energy} (2$\times10^{-7}$ M$_\odot$ yr$^{-1}$).  In general, mass loss rates are expected to be about 10\% that of the accretion rate. The disk itself has a mass of order 10$^{-2}$ M$_\odot$ \citep{Tilling12}, which, when coupled with the mass loss rate we derive, suggests a disk dispersal timescale of 10$^5$ yr. Given the age of the system (4 Myr), it is possible that the star is in the final stages of accretion, and is blowing out the last of its infalling material via the final stages of a disk wind.  We note that \citet{Kraus08} suggest that the Br$\gamma$ accretion rate for this source may be contaminated by outflow emission, which further strengthens our argument, as the mass loss rate would then be higher than the mass accretion rate. If the gas were from a photo evaporative disk wind \citep[i.e.][]{HollenbachPPIV}, then the characteristic velocity of the gas measured here (-18.6 km s$^{-1}$) should be roughly equivalent to the sound speed.  A sound speed of 18.6 km s$^{-1}$ indicates a temperature of 10$^5$ K, which is much too high for molecular gas to be present. Thus, we are not seeing a photo evaporative wind.

\begin{table}
\caption{Outflow energetics for the two CO transitions.  The first CO $J$=2-1 column is derived from integrating over all of the CO $J$=2-1 emission, while the second is derived from the data that was filtered to sample the same spatial scales as the Band 7 data. This `Filtered' column is more directly comparable to the $J$=3-2 data.  The quoted uncertainties come from the uncertainties in the fluxes, propagated through the equations in quadrature. Note that the quoted velocity is blue shifted with respect to the source velocity of 5.8 km s$^{-1}$. }
\begin{tabular}{lr@{$\pm$}lr@{$\pm$}lr@{$\pm$}l}
\hline
\hline
 & \multicolumn{2}{c}{CO $J$=3-2} & \multicolumn{2}{c}{CO $J$=2-1}&\multicolumn{2}{c}{ CO $J$=2-1}\\
 &\multicolumn{2}{c}{}&\multicolumn{2}{c}{Integrated} & \multicolumn{2}{c}{ Filtered}\\
 \hline
M (10$^{-6}$ M$_\odot$)&	5.7	&	0.06	&	73.1	&	0.8	&24.6 &0.3\\
V (km s$^{-1}$) &		-18.6	&	0.1	&	-18.6	&	0.3	&-18.9&0.5\\
P	(10$^{-5}$M$_\odot$ km s$^{-1}$)&	10.7	&	0.1	&	136.0	&	3.5& 46.4&4.6	\\
E (10$^{40}$ erg)& 		1.98	&	0.04	&	25.1	&	1.0&	8.7&0.5\\
\hline
\multicolumn{7}{c}{Wind kinematic age = 400 yr}\\
\hline
 F$^a$  & 2.7&0.3 &34.0&4 & 11.5 & 2\\
L (10$^{-3}$ L$_\odot$)& 		0.41	&	0.05	&	5.2	&	0.7	&1.8&0.3\\
$\dot{\rm M}_{\rm out}$ (10$^{-8}$M$_\odot$ yr$^{-1}$)	&	1.4	&	0.2	&	18.3	&	2&6.2&0.7	\\
 \hline
 \hline
\multicolumn{7}{p{0.8\columnwidth}}{ $^a$ The units of the wind force are  \mbox{10$^{-7}$ M$_\odot$ km s$^{-1}$ yr$^{-1}$}.}
\end{tabular}
\label{tab:energy}
\end{table}

\section{Conclusions}
\label{sec:conclusions}

We have presented here the first evidence for a molecular disk wind coming from a Herbig Ae star. This disk wind is  coincident with a chain of HH knots.   We have quantified the energetics of the blue shifted wind emission, and show that the positions of the two CO $J$=3-2 emission knots are consistent with one of the optically detected HH objects (A2), and a younger knot (A3) which was likely emitted during a flare in 2002.  Further followup ALMA observations, including wider-field mosaicing and incorporating larger scale emission will be required to fully characterize this phenomena.

 The mass being released from the system via the disk wind is twice  the mass accretion rates found via Br$\gamma$ emission \citep{Garcia06}. For stable, accreting systems, this ratio is generally closer to 0.1 \citep[e.g.][]{Hartigan95}.  The derivation of the accretion rate using Br$\gamma$ is a time average \citep{Muzerolle98}, and since the age used for the wind is kinematically derived, so is the mass loss rate in the wind.  Because the mass loss rate is, on average, larger than the mass accretion rate, this system cannot be in a stable growth stage.

\begin{acknowledgements}
This paper makes use of the following ALMA data: ADS/JAO.ALMA\#2011.0.00010.SV. ALMA is a partnership of ESO (representing its member states), NSF (USA) and NINS (Japan), together with NRC (Canada) and NSC and ASIAA (Taiwan), in cooperation with the Republic of Chile. The Joint ALMA Observatory is operated by ESO, AUI/NRAO and NAOJ.  IdG-M acknowledges the Spanish MICINN grant AYA2011-30228-C03 (co-funded with FEDER funds). AH acknowledges support from Millennium Science Initiative, Chilean Ministry of Economy: Nucleus P10-022-F. Astrochemistry in Leiden is supported by NOVA, KNAW and EU A-ERC grant 291141 CHEMPLAN. Allegro is funded by NWO Physical Sciences ("the Netherlands Organization for Scientific Research (NWO), Physical Sciences")
\end{acknowledgements}

\bibstyle{aa}

\end{document}